\documentclass[conference]{IEEEtran}

\usepackage{cite}
\usepackage{amsmath,amssymb,amsfonts}
\usepackage{algorithmic}
\usepackage{comment}
\usepackage{graphicx}
\graphicspath{{Images/}}
\usepackage{tabularx}
\usepackage{textcomp}
\usepackage{xcolor}
\usepackage{tikz}
\usepackage{siunitx}
\usetikzlibrary{shapes.geometric,arrows}
\tikzstyle{process}=[rectangle, minimum width=3cm, minimum height=.5cm, text centered, draw=black, fill=orange!30 ]{\small}
\tikzstyle{process2}=[rectangle, minimum width=1 cm, minimum height=.5cm, text centered, draw=black, fill=white ]{\footnotesize}
\tikzstyle{arrow} = [thick,->,>=stealth]
\usepackage[scaled=.7]{beramono}
\usepackage{microtype}
\newcommand{\para}[1]{\vspace{3pt}\noindent\textbf{#1.}}
\usepackage{graphics}
\usepackage{color}

\begin{document}

\title{On the Feasibility of Exploiting Traffic Collision Avoidance System Vulnerabilities}


\author{\IEEEauthorblockN{Paul M. Berges}
\IEEEauthorblockA{\textit{Virginia Tech}
\\pberges7@vt.edu
}
\and
\IEEEauthorblockN{Basavesh Ammanaghatta Shivakumar}
\IEEEauthorblockA{\textit{Purdue University}
\\bammanag@purdue.edu
}
\and
\IEEEauthorblockN{Timothy Graziano}
\IEEEauthorblockA{\textit{Virginia Tech}
\\tmgraz@vt.edu
}
\and
\IEEEauthorblockN{Ryan Gerdes}
\IEEEauthorblockA{\textit{Virginia Tech}
\\rgerdes@vt.edu 
}
\and
\IEEEauthorblockN{Z. Berkay Celik}
\IEEEauthorblockA{\textit{Purdue University}
\\zcelik@purdue.edu
}
}
\maketitle

\begin{abstract}
Traffic Collision Avoidance Systems (TCAS) are safety-critical systems required on most commercial aircrafts in service today. However, TCAS was not designed to account for malicious actors. While in the past it may have been infeasible for an attacker to craft radio signals to mimic TCAS signals, attackers today have access to open-source digital signal processing software, like GNU Radio, and inexpensive software defined radios (SDR) that enable the transmission of spurious TCAS messages. In this paper, methods, both qualitative and quantitative, for analyzing TCAS from an adversarial perspective are presented. To demonstrate the feasibility of inducing near mid-air collisions between current day TCAS-equipped aircraft, an experimental \textit{Phantom Aircraft} generator is developed using GNU Radio and an SDR against a realistic threat model.
\end{abstract}


\section{Introduction}
Aviation remains the safest way to travel because of the various safety-critical systems operating at any given moment on the aircraft~\cite{misc:safest_ways_to_travel}.  One such on-board safety feature is known as the Traffic Collision Avoidance System (TCAS), internationally known as the Airborne Collision Avoidance System (ACAS), that prevents the mid-air collisions of transponder-equipped aircraft.  In the event of a Near Mid-air Collision (NMAC) where Air Traffic Control (ATC) towers cannot react in time, TCAS is critical for warning pilots to change course and prevent a mid-air collision.  Many aviation regulatory bodies mandate the use of TCAS on larger commercial aircraft~\cite{misc:acas_ii_mandate_eu}. 

Recent technology, such as Software Defined Radio (SDR), enables the manipulation of TCAS through software-defined wireless signals designed to appear like one or more aircraft on a collision course with a target aircraft. TCAS was never intended to perform under adversarial conditions which are entirely feasible for a malicious actor to create in today's environment of inexpensive, powerful computers.  If an attacker were to compromise TCAS, they could bypass the safety benefits granted by TCAS equipage. Worse, under certain conditions, a TCAS-equipped aircraft could have a higher chance of mid-air collision than an unequipped aircraft.  

Most prior work into the security of the Mode S transponder has been limited to Automated Dependent Surveillance-Broadcast (ADS-B) message spoofing~\cite{proc:lhcsas, proc:sat} or pilot responses to erroneous TCAS messages but does not specify the technical requirements to spoof TCAS~\cite{misc:pilot_response}. This paper represents the first research into accurately spoofing TCAS messages, and the danger of induced near mid-air collisions as a result of the aforementioned spoofed messages. TCAS and ADS-B are closely linked because both messages are transmitted through the Mode S (Selective Aeronautical telecommunication; able to be interrogated) transponder. While works on ADS-B, propose some defenses against spoofing attacks, ADS-B is not part of any safety-critical systems on an aircraft, negating the need for sweeping and expensive security changes. If TCAS is also shown to be vulnerable to attack, then the safety of onboard passengers would be compromised, providing a more significant motivator to institute system changes and improvements. Due to the safety implications of this research, our work was originally done in conjunction with a primary manufacturer of TCAS and with notification to Department of Homeland Security (DHS).  

In this paper, we demonstrate that the TCAS transactions are vulnerable to attack.  We explore TCAS vulnerabilities using the weakest possible attack model using only open-source software, publicly-accessible knowledge about TCAS, and a low cost SDR.  We hope that our explorations motivate further research into the defense of TCAS and Mode S transponders. In this work, we make the following contributions:
\begin{itemize}
\item We show that a safety-focused design approach for safety-critical systems does not result in a safe system in an adversarial environment. 
\item We take an analytical approach to find vulnerabilities in, and explore the effect of failures of, TCAS II through an attack tree.
\item We develop a Phantom Aircraft attack, where we use open-source software to spoof critical TCAS II components (e.g., a Mode S Transponder) enabling an attacker to masquerade as a collision-bound aircraft.
\end{itemize}

\section{Background}
We provide background on TCAS and methods for attack construction to aid understanding and evaluation of the Phantom Aircraft attack provided in the subsequent sections.

\subsection{Traffic Collision Avoidance System (TCAS)}
\begin{figure}[t!]
    \centering
    \includegraphics[width=0.85\columnwidth]{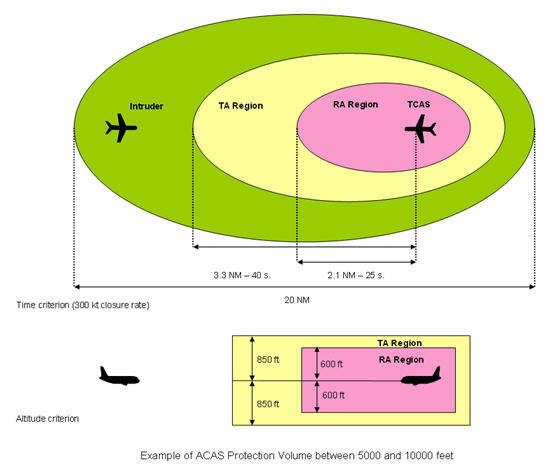}
    \caption{The region protected by TCAS~\cite{tech:skybrary_acas}.}
    \label{fig:tcasProtection}
\end{figure}

TCAS is designed to reduce the risk of Near Mid-air Collisions (NMAC) between aircraft. It operates independently of ATC to notify/instruct pilots when a protected volume of airspace around the aircraft is intruded upon by other aircraft. It uses an onboard transponder (transmitter/receiver) to facilitate air-to-air communications between aircraft. Each TCAS interrogates nearby aircraft on the \SI{1030}{\mega\hertz} frequency band and all aircraft reply on the \SI{1090}{\mega\hertz} frequency band. TCAS uses these transactions to track nearby aircraft and monitor for intrusions in nearby airspace. Figure~\ref{fig:tcasProtection} shows the protected region generated by TCAS defined by altitude and the remaining time until an intruder reaches its Closest Point of Approach (CPA)~\cite{tech:skybrary_acas, manual:intro_tcas_ii}. TCAS is designed to operate independently of  ATC and other tracking systems in order to mitigate collision risk regardless of the operational state of other systems or ground based information relayed to the pilot. 

\begin{figure}[t]
    \centering
    \includegraphics[width=0.5\columnwidth]{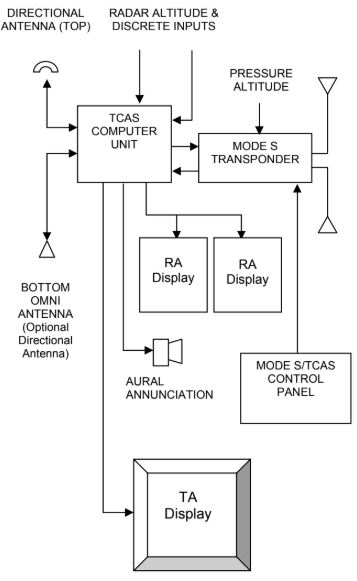}
    \caption{TCAS System Components ~\cite{manual:intro_tcas_ii}}
    \label{fig:tcasSystemComponents}
\end{figure}

\para{TCAS System Components} TCAS requires that aircraft be equipped with a transmitter/receiver known as a Mode S transponder.  TCAS itself consists of three main components illustrated in Figure~\ref{fig:tcasSystemComponents}:
\begin{itemize}
    \item \textbf{TCAS CPU:} A computer that performs interrogations, surveillance, tracking, threat declaration, etc.
    \item \textbf{Antennas:} There are three required antennas and one optional antenna. However, many implementations choose a shared antenna design to save costs.
    \item \textbf{Cockpit displays:} Information on intruding aircraft and suggested responses are visually and aurally announced to pilots.
\end{itemize}

\para{TCAS: Request-Response} Generally, TCAS behaves within a request-response architecture. A TCAS CPU uses the \SI{1030}{\mega\hertz} channel to request information from an intruder. This action is called an \textit{interrogation}. The Mode S transponder on an intruding aircraft is responsible for responding to interrogations~\cite{manual:acas_manual, tech:ssr_mode_s_advisory}. Each plane has a unique 24-bit address called its International Civil Aviation Organization (ICAO) address~\cite{manual:acas_manual, manual:annex10_vol4}. All interrogations are addressed to the aircraft receiving the interrogation. ICAO addresses are combined using an exclusive-or function with a unique Mode S cyclic redundancy check (CRC) calculation to form the Address/Parity (AP) field~\cite{manual:modes_crc_calc}. The data encoded in interrogation is referred to as an Uplink Format (UF) message. UF messages have a fixed-length 5-bit header, and its value indicates what type of UF message it is. Replies are broadcast on the \SI{1090}{\mega\hertz} frequency band and only happen if an interrogation’s decoded AP field matches the ICAO address of the receiving aircraft. The data encoded in a reply is referred to as a Downlink Format (DF) message. The DF header’s value will always match the UF header’s value of the message that started the transaction. 

\para{TCAS advisories}
There are two types of advisories (i.e., responses to intrusions) that TCAS can produce.
\begin{itemize}
    \item A \textbf{Traffic Advisory (TA)} is intended to help a pilot visually locate an intruding aircraft; it is issued first.
    \item A \textbf{Resolution Advisory (RA)} will recommend maneuvers or positional holds in order to increase the separation between the aircraft and intruder~\cite{manual:intro_tcas_ii}.
\end{itemize}

In a coordinated maneuver, the aircraft that announces its RA first controls the maneuvers of the second aircraft. In other words, the aircraft that makes the first decision will coordinate the possible maneuvers the other aircraft can do. In the event of simultaneously received RA coordination, the lower ICAO identification number receives precedence. An attacker can control this within the spoofed message to always assert primacy. Then, the TCAS will choose the maneuver’s \textit{sense} and \textit{strength}~\cite{manual:intro_tcas_ii}. A sense is the direction the aircraft will maneuver. After a sense selection, TCAS will choose a maneuver strength. A strength can either be positive or negative. Upon receipt of a RA, the pilot is obliged to follow the RA maneuver guidance if possible, giving higher priority to the RA than ATC guidance. Related work has shown that pilots will follow an RA even if they believe the system to be behaving inappropriately or ``crying wolf''~\cite{misc:pilot_response}. 

\para{TCAS Operating Modes}
TCAS can be operated in three modes, which are controlled by the pilot~\cite{manual:intro_tcas_ii}. The level of functionality TCAS provides is defined by its Sensitivity Level (SL). These levels are:
\begin{itemize}
    \item (SL 1) \textbf{Standby}: No TCAS tracking or ``squitters'' (a non-solicited Mode S transmission of aircraft tracking data to alert nearby transponders of the aircraft's presence). Mode S transponder will respond to interrogations.
    \item (SL 2) \textbf{TA Only}: TCAS will track intruding aircraft, but it will only announce a TA to the pilot. Mode S transponder transmits squitters and responds to interrogations.
    \item (SL 3+) \textbf{TA-RA}: TCAS will automatically select the best SL for the altitude. 
\end{itemize}

In some instances ATC towers can partially control TCAS. Specifically, ground controllers can place TCAS into SL 2 or greater~\cite{manual:intro_tcas_ii, manual:annex10_vol4, manual:acas_manual}. Only after repeated TCAS malfunctions will a pilot degrade the SL~\cite{misc:pilot_response}.

\subsection{Software-defined Radio}
A Software-defined radio (SDR) does not use dedicated hardware circuits for signal processing but instead conducts radio-relevant processing in software providing flexibility in terms of modulation, filtering, operating frequency, and frame format~\cite{mitola1995software}. GNU Radio is a free and open-source development platform for SDR that uses a graphical approach to radio design and supports development in C++ or Python~\cite{misc:gnuradio}. It is often used in a simulation environment to support real-world radio systems and wireless communications research~\cite{blossom2004exploring}. GNU Radio Companion (GRC) is a graphical application packaged with GNU Radio~\cite{misc:gnuradio}.

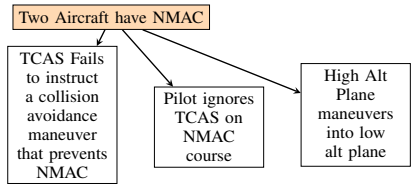
\begin{figure}[t!]
    \centering
    \scalebox{.65}{
    \begin{tikzpicture}[node distance=2cm]
    \node (pro1) [process, xshift=4cm] {Two Aircraft have NMAC};
    \node (pro2) [process2, text width=2cm, below of=pro1, xshift=-1cm] {TCAS Fails to instruct a collision avoidance maneuver that prevents NMAC};
    \node (pro3) [process2, text width=2cm, right of=pro2, xshift=1cm,yshift=-.25cm] {Pilot ignores TCAS on NMAC course};
    \node (pro4) [process2, text width=2cm,right of=pro3, xshift=1cm,yshift=.25cm] {High Alt Plane maneuvers into low alt plane};
    \draw [arrow] (pro1) --  (pro2);
    \draw [arrow] (pro1) --  (pro3);
    \draw [arrow] (pro1) --  (pro4);
    \end{tikzpicture}
    }
    \caption{The root and major leaf nodes of the TCAS attack tree}
    \label{fig:atktree_top}
\end{figure}

\subsection{Attack Tree and Fault Tree Analysis}
An attack tree is a graphical representation of an attack including simply a series of “What if ...” questions that result is an extensive list of known attack vectors to the system~\cite{misc:schneier_atk_trees}. Fault Tree Analysis (FTA) is a tool that safety engineers use to inform decision making concerning design revision with respect to some undesired event~\cite{tech:fta_overview}. FTAs are created to model failure in very complex systems and can show how a series of events can lead to a larger undesired failure event using probability, Boolean algebra, etc. Similar to the attack tree, the root of an FTA starts from a goal (i.e., starting from the desired failure to investigate) and then deduces the specific components that could cause the failure.

\section{TCAS Attack Tree}
\label{sec:attack_tree}
We consider an attacker who aims to induce an NMAC. The attack tree of this work shares two major components with the TCAS fault tree; the major components identified are shown as the top of the attack tree in Figure \ref{fig:atktree_top}.  The left event represents \emph{unresolved NMAC} attacks (e.g.,  two planes are already on a collision course, and the pilots fail to maneuver the planes in a way that avoids NMAC), and the right event represents the \emph{induced NMAC} attacks (e.g., two planes that were not previously on course towards NMAC are maneuvered into an NMAC). The central event accounts for any attacks on people like social engineering~\cite{manual:social_engineering}. Our particular interest is the induced component of NMAC because it is antithetical to the design goals of TCAS and implies that an attacker could intentionally redirect two nearby planes into each other.

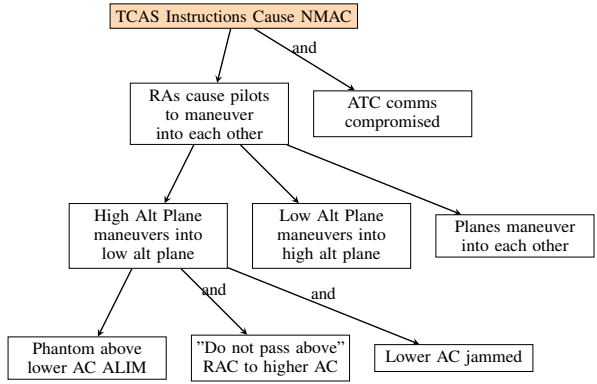
\begin{figure}[t!]
    \centering
    \scalebox{.65}{
\begin{tikzpicture}[node distance=2cm]
\node (pro1) [process, xshift=4cm] {TCAS Instructions Cause NMAC};
\node (pro2) [process2, text width=3cm, below of=pro1, xshift=-.5cm] {RAs cause pilots to maneuver into each other};
\node (pro3) [process2, text width=3cm, right of=pro2, xshift=1.75cm,] {ATC comms compromised};
\node (pro4) [process2, text width=3cm, below of=pro2,yshift=-.5cm, xshift=-1.25cm,] {High Alt Plane maneuvers into low alt plane};
\node (pro5) [process2, text width=3cm, right of=pro4, xshift=1.75cm,] {Low Alt Plane maneuvers into high alt plane};
\node (pro6) [process2, text width=3cm, right of=pro5, xshift=1.75cm,] {Planes maneuver into each other};
\node (pro7)[process2, text width=3cm, below of=pro4,yshift=-.5cm, xshift=-1.25cm,] {Phantom above lower AC ALIM};
\node (pro8)[process2, text width=3cm, right of=pro7, xshift=1.75cm,] {"Do not pass above" RAC to higher AC};
\node (pro9) [process2, text width=3cm, right of=pro8, xshift=1.75cm,] {Lower AC jammed};
\draw [arrow] (pro1) --  (pro2);
\draw [arrow] (pro1) --  node[anchor=south] {and}(pro3);
\draw [arrow] (pro2) --  (pro4);
\draw [arrow] (pro2) --  (pro5);
\draw [arrow] (pro2) --   (pro6);
\draw [arrow] (pro4) --   (pro7);
\draw [arrow] (pro4) -- node[anchor=south] {and}(pro8);
\draw [arrow] (pro4) -- node[anchor=south] {and}(pro9);
\end{tikzpicture}
}
    \caption{A simplified version of the branch on induced NMAC attacks.}
    \label{fig:atktreeInduc}
\end{figure}
\para{Induced NMAC Attacks}
Induced NMAC attacks are reliant on fooling TCAS into generating an RA that, if followed, would induce an NMAC. A section of the induced NMAC attack is shown in Figure~\ref{fig:atktreeInduc}.  Attacks that would induce an NMAC require that the expected Mode S messages are obeyed to trick a TCAS into a \textit{false track}, i.e., following a non-existent \emph{Phantom Aircraft} course, whose purpose is to cause the target aircraft to maneuver onto a probable collision course with another aircraft. TCAS implicitly trusts that responses to its interrogations originate from an actual aircraft and that the aircraft will travel in a certain manner. If an attacker can respond to TCAS's interrogations and appear to move like a plane, then the false track should be successfully created and maintained.  The foundation of the Phantom Aircraft starts with an accurate spoofer of Mode S signals; it must also consistently fool the range measurements that TCAS performs~\cite{manual:acas_manual, proc:rfid_army_knife}.   The induced NMAC attacks are reliant on two threats being near in altitude to increase the probability of a successful attack. 
\section{Phantom Aircraft Attack}
We develop a particular induced NMAC attack: the Phantom Aircraft attack.  First, we define the threat model and outline the attack model. This represents a specific example of a TCAS attack. The design of any attack in the attack tree of Section~\ref{sec:attack_tree} would follow similar steps to the procedure shown.

\para{Threat Model}
The attacker's goal is to induce an NMAC between two aircraft without any modification to the TCAS on the aircraft. The attack therefore requires replying to relevant interrogations as well as emulating \textit{distance closing} signal behavior, in which the Round Trip Time (RTT) range estimation of TCAS is fooled with proper response timing into believing that an aircraft is intruding. The attacker launches the attack using GNU Radio~\cite{misc:gnuradio} and a relatively low-cost SDR hardware from Ettus Research~\cite{misc:ettus}. In a short list, the attacker's capabilities are the following:
\begin{itemize}
    \item The attacker is equipped with an SDR and directional antennas capable of transmitting signals to the targeted aircraft.
    \item The location and speed of the targeted aircraft are known to the attacker.
    \item The attacker is able to selectively jam Mode S transmissions of aircraft using knowledge of their ICAO address and its reply periodicity. Directional jamming is not required given this knowledge.
    \item The attacker can properly time their replies, with respect to the interrogation periodicity of an aircraft, to appear closer or farther than they really are.
\end{itemize}

\begin{figure}[t!]
    \centering
    \includegraphics[width=1\columnwidth]{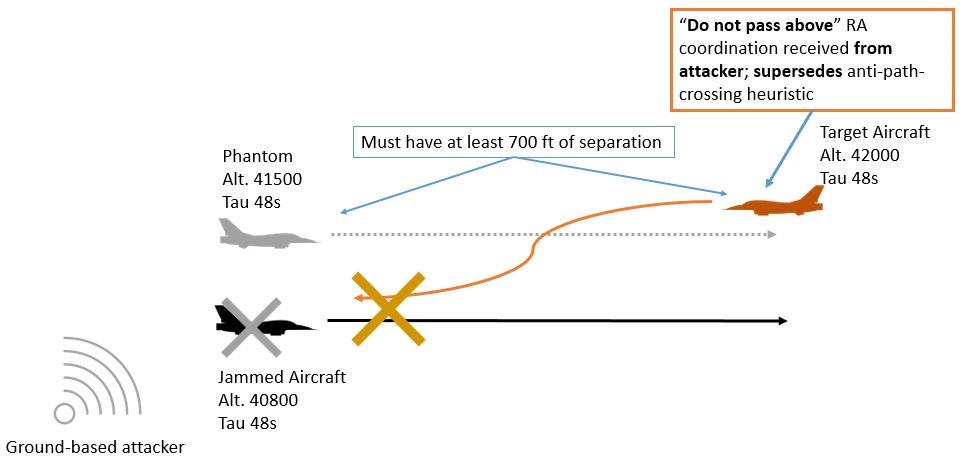}
    \caption{Planned Phantom Aircraft attack against two nearby intruders}
    \label{fig:atkdesign_encounter}
\end{figure}

\para{Attack Model} The intended encounter is the scenario shown in Figure~\ref{fig:atkdesign_encounter}. For analysis two aircraft are approaching each other without any horizontal offset. These aircraft are also approaching head-on such that there is an exact 180-degree difference between their bearings with no vertical rate of approach. Alternate approach angles do not impact the feasibility of the attack. The aircraft are placed such that they have the adequate vertical separation that would not elicit any TCAS warnings. The attacker begins selectively jamming the black aircraft (the lower one on the left in Figure~\ref{fig:atkdesign_encounter}) to drop its track from the target aircraft. A false track is baited, and maintained, by the attacker on the target aircraft. By analyzing the expected RTT of the request-response TCAS handshake, an attacker can manipulate artificial latency by delaying their reply once the interrogation is received, in order to appear to be at a desired distance and closure rate. This calculation requires knowledge of the round trip distance between the victim and phantom times the speed of light, factoring the computational latency of a TCAS module upon message receipt, and the ability for the attacker to transmit spoofed messages with \SI{}{\nano\second} timing resolution.

Once the attacker has determined that the target is within the TA range of the phantom, the phantom must send its Resolution Advisory Complement (RAC) first. The attacker can simply announce an RA arbitrarily because TCAS trusts that the RAC it receives is from a transponder acting in good faith. The attacker forces the target TCAS to cross paths with the phantom by constructing a ``Do not pass above" coordination~\cite{manual:annex10_vol4}. In combination with the required Altitude Limit (ALIM) for this flight level, the target would need to descend to an altitude no greater than 40,800 ft. 

To accomplish the Phantom Aircraft attack, five conditions must be met. The attacker needs to:
\begin{enumerate}
    \item Perform reconnaissance by interrogating the airspace in the attacker's vicinity.
    \item Estimate the trajectory of victim aircraft through tracking.
    \item Bait a false track from the victim by emitting squitters and detecting them when interrogations begin.
    \item Maintain the false track by responding to interrogations.
    \item Declare a threat against the victim and detect evidence that the victim declared its own RA.
\end{enumerate}

\section{Attack Implementation}
\label{sec:attack-implementation}
We developed an attack platform based on GNU Radio.  We present the block diagrams of the platform and simulation methods. Lastly, an outline of the hardware used for this platform is presented.

\subsection{Phantom Aircraft Generator}
\begin{figure}[t!]
    \centering
    \includegraphics[width=1\columnwidth, keepaspectratio]{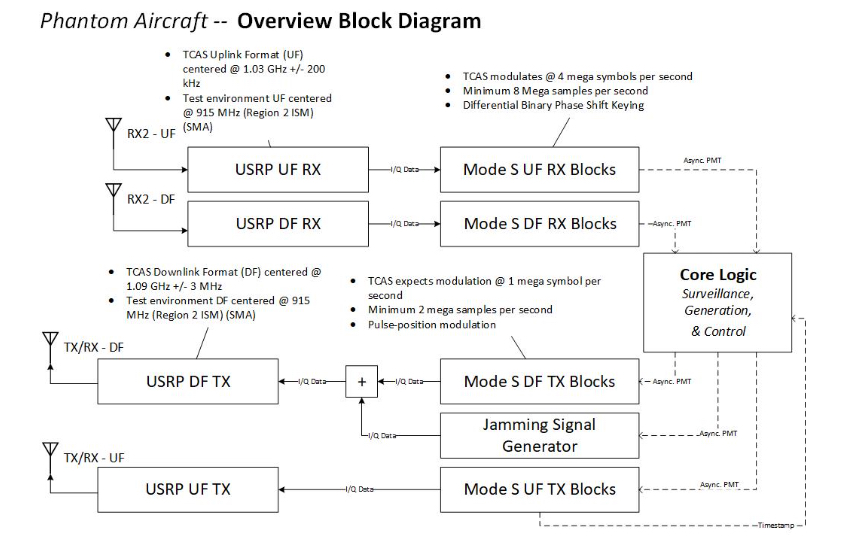}
    \caption{Overview block diagram of the phantom aircraft generator.}
    \label{fig:blockDiagramOverview}
\end{figure}

We demonstrate a proof-of-concept phantom aircraft generator for GNU Radio application to bait a TCAS into tracking a phantom aircraft by having a pair of zero-speed aircraft track each other. The core functionality of an attacker and a normal aircraft are identical. Figure~\ref{fig:blockDiagramOverview} shows the overview of the phantom aircraft generator as well as expected carrier frequencies, modulations, and sample rates.

We designed the system using open-source GNU Radio modules. The reception and demodulation of DF packets are handled by \textit{gr-adsb} (ADS-B OOT for GNU Radio)~\cite{misc:gh_adsb}. The total OOT dependencies for gr-modes are: (1) \texttt{gr-adsb}: demodulation and decoding of ADS-B messages~\cite{misc:gh_adsb}, (2) \texttt{gr-burst}: blocks for building PSK modems~\cite{misc:gh_burst}, (3) \texttt{gr-eventstream}~\cite{misc:gh_eventstream}, (4) \texttt{gr-mapper}~\cite{misc:gh_mapper}, and (5) \texttt{gr-pyqt}~\cite{misc:gh_pyqt}.

Universal Software Radio Peripheral (USRP) hardware requires significant processing power in general, especially at high sample rates.  The PC for the attack uses an Intel Core i7-6800K; six cores at a 3.4 GHz clock rate with 16GB RAM. An Ettus Research B210 is the transmitting USRP, and an Ettus Research N210 is the receiving USRP~\cite{manual:b210, manual:n210}. These USRPs are low cost, \$1,259 and \$2,011 respectively. The two USRPs communicate over SMA cables through a variable attenuator that prevents the receiver's analog to digital converter (ADC) from being saturated.  

\subsection{Phantom Aircraft Attack Details}
We present the architecture of the phantom aircraft generator, the sub-routines of each high-level block, and the interfaces between each function. 

Our system uses both asynchronous and streamed data. A TX and RX interface for both interrogations and replies is used. A nearby carrier frequency in the industrial, scientific, and medical (ISM) band is chosen for testing~\cite{manual:ism_bands} instead of normal TCAS frequency bands~\cite{manual:annex10_vol4, manual:acas_manual, manual:intro_tcas_ii}. Each RX interface converts raw samples into packets the \texttt{Core Logic} can understand; each TX interface performs the opposite. The range estimation occurs from the RTT of the interrogation-reply cycle using timestamp feedback from the TX and RX blocks.  

Conversion of data from the \texttt{Core Logic} to samples interface also handles the packing of message data into a complete Mode S message. A \texttt{lambda function} block from \texttt{gr-pyqt} and a \texttt{prepend preamble} block from \texttt{gr-burst} are convenient for packing the data into a Mode S reply~\cite{misc:gh_pyqt, misc:gh_burst}.  

The data that enters the DF TX interface from the \texttt{Core Logic} is 56 or 112 samples of bytes in length. To convert this message into a Mode S reply, the following procedure is performed: (a) Pulse Position Modulation (PPM) is performed~\cite{art:ppm}, (b) the Mode S reply preamble is attached to the beginning of the packet, and (c) the packet is interpolated to the application's base sample rate. Then, the packet must be converted to a \textit{stream} of complex samples the GNU Radio supports. Once the payload is packed into a complete Mode S reply, \texttt{gr-eventstream} is used to insert the data from an \textit{asynchronous} packet into the sample stream.

The receiver interface for DF messages is made from \texttt{gr-adsb} blocks~\cite{misc:gh_adsb}. The \texttt{ADS-B Framer} is slightly modified such that it creates a timestamp that propagates to the \texttt{Core Logic} when a preamble is detected.  The \texttt{ADS-B Framer} block correlates the input with the expected preamble and creates a \texttt{tag} object on the first sample of the preamble when a match is detected.  This \texttt{tag} serves as the synchronization point which the \texttt{ADS-B Demodulator} can use for PPM demodulation. We mimic the design of the reply transmission interface such that the interface remains independent of actual message data. Some adjustments are made to support the different pulse sequence and modulation scheme of Mode S interrogations and to suppress Modes A and C transponders from processing the interrogation.  

The receiving interface for the Interrogation Framer block performs correlation analysis of the preamble and creates a \texttt{tag} on the first sample of the preamble. Using this as a synchronization point, the \texttt{Tag Consumer} block then creates a ``slice" of samples from 1-120 Differential Binary Phased Shift Keyed (DBPSK) samples (i.e., length of the preamble + 112 symbols assumed). Demodulation is performed asynchronously. The \texttt{Core Logic} interfaces then convert interrogation and reply samples into a format the state machine logic can understand.

\section{Evaluation}
We verify the attack detailed in Section~\ref{sec:attack-implementation}. While the direct testing approach would be to generate aircraft on a real TCAS, real TCAS hardware is not trivial to acquire and setup. Therefore, verification is completed with two simplified aircraft cores that can track each other at zero speed through the medium of Mode S messages. We performed a series of unit and integration tests through GNU Radio to verify the adversarial tasks. For each point of verification below, we executed the flowgraph for 10 minutes with a fixed surveillance period of one second such that approximately 600 rounds of interrogations and replies occur for each item. We have verified the following:
\begin{enumerate}
    \item Arbitrary Mode S reply messages can be crafted to fool the \texttt{gr-adsb} receiver without USRP hardware in-the-loop~\cite{misc:gh_adsb}).
    
    \item The \texttt{gr-adsb} receiver's performance is characterized via a packet-loss calculation with USRP hardware-in-the-loop.
    
    \item Arbitrary Mode S interrogation messages can be crafted to fool a custom Mode S interrogation receiver without USRP hardware-in-the-loop. 
    
    \item All four message modulators and demodulators can deliver payloads to the \texttt{core logic}. The core logic performance is indicative of the state machine functionality. No USRP hardware is used in-the-loop.
\end{enumerate}
\begin{figure}[ht!]
    \centering
    \includegraphics[width=1\columnwidth]{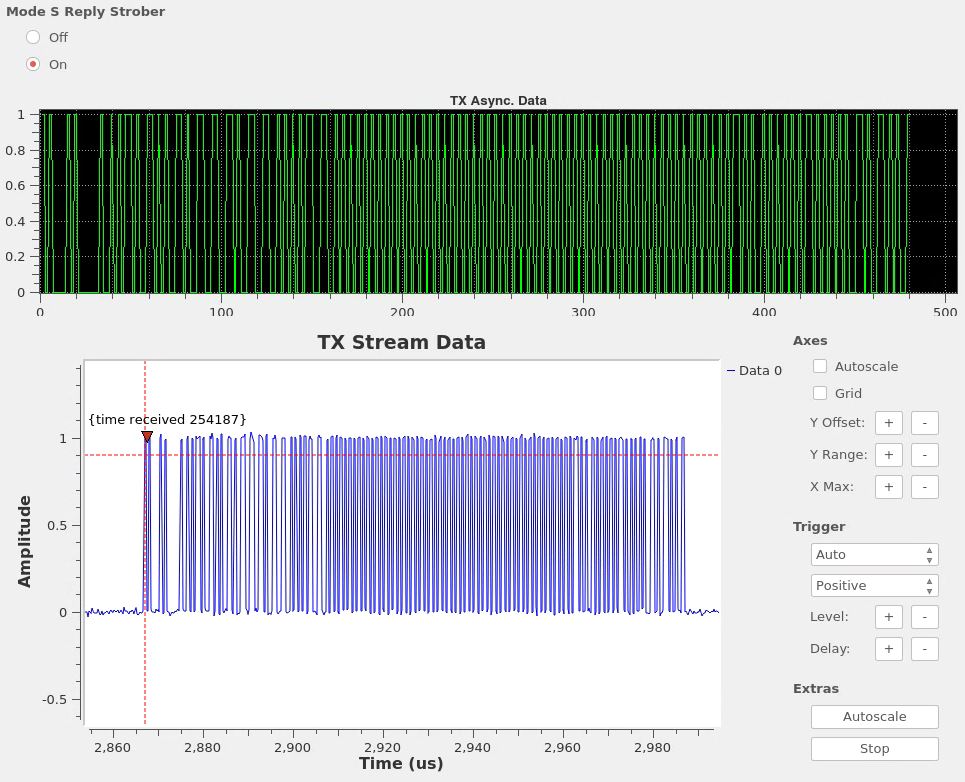}
    \caption{Real-time graph showing the DF 17 message received and decoded.}
    \label{fig:verify_df_txrx_nohw_gui}
\end{figure}

\para{Mode S Reply}
We first verified that we correctly modulated, demodulated, and decoded a crafted payload message as shown in Figure~\ref{fig:verify_df_txrx_nohw_gui}. Six different 10-minute test runs are performed. We found that even the best performing case still has a packet-loss factor nearly 60\%. While this is a significant problem for hardware testing, our primary focus is its emulation of TCAS functionality. Therefore, the remainder of the tests ignores the use of USRP hardware. 
%
\begin{figure}[t!]
    \centering
    \includegraphics[width=1\columnwidth]{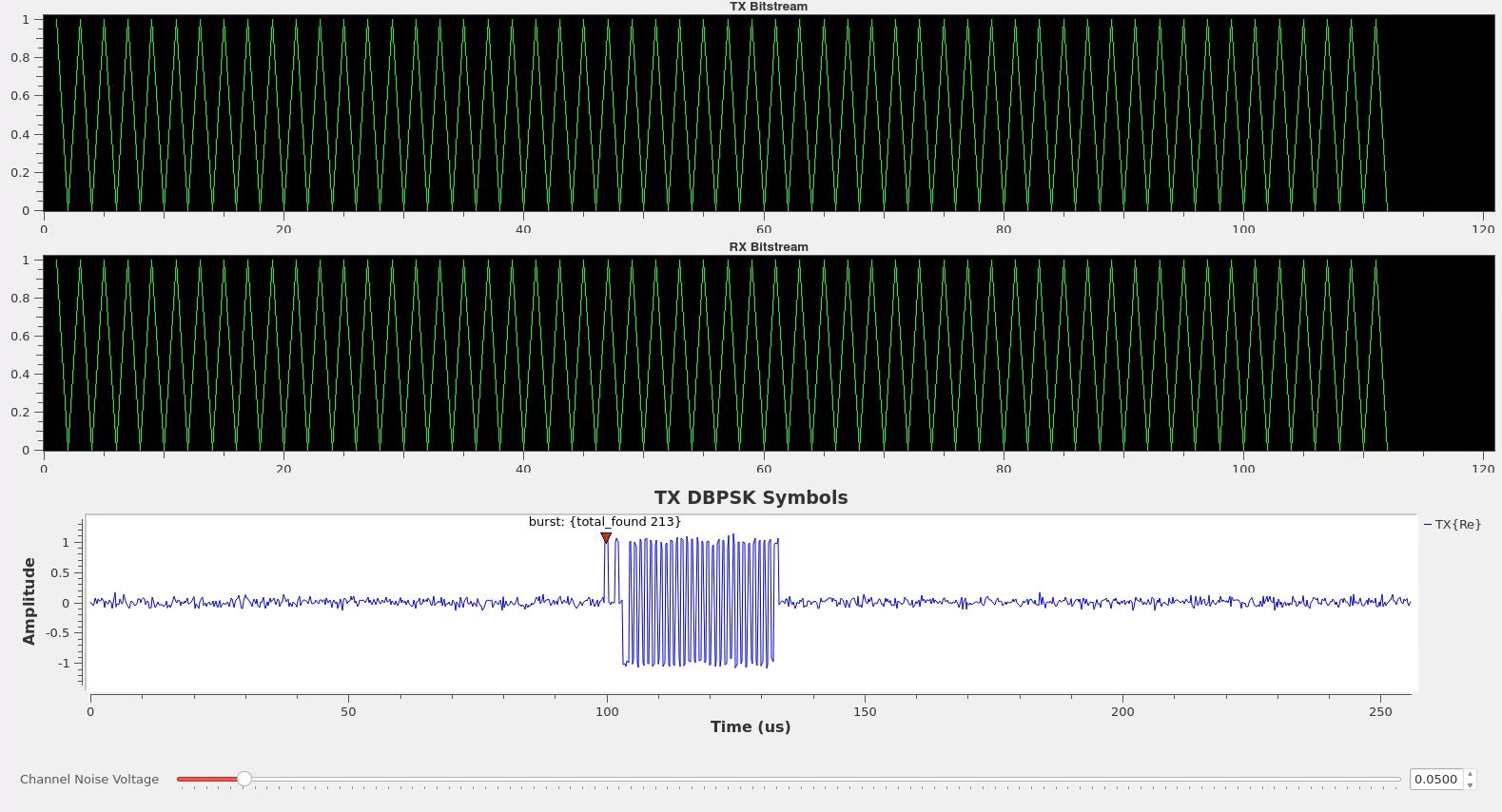}
    \caption{Three real-time plots of the interrogation interface test.  The top plot shows the asynchronous TX bit stream, the middle plot shows the asynchronous RX bit stream, and the lower plot shows the real-time plot of the \texttt{framer} marking the start of the message.}
    \label{fig:verify_uf_txrx_nohw_gui}
\end{figure}

\para{Mode S interrogation} We verified that an arbitrary payload of 112 bits is modulated and demodulated to confirm an expected Mode S interrogation through a simulated Additive White Gaussian Noise (AWGN) channel as shown in Figure~\ref{fig:verify_uf_txrx_nohw_gui}. 

\begin{figure}[t!]
    \centering
    \includegraphics[width=1\columnwidth]{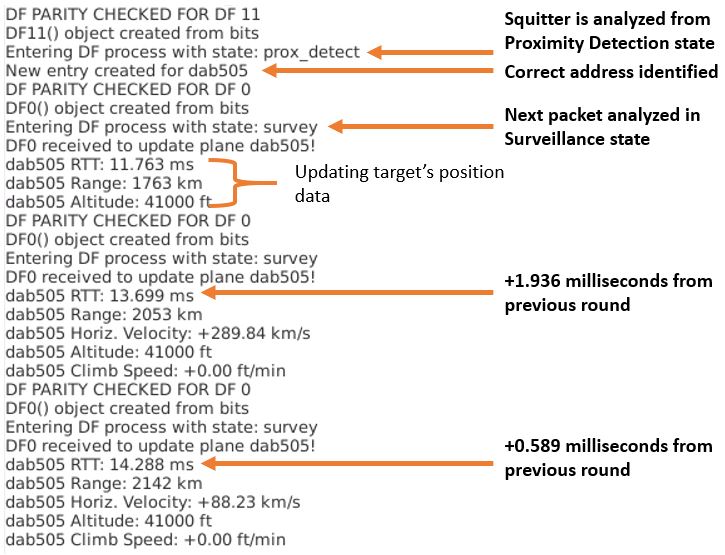}
    \caption{Imprecise RTT measurements from the design under test}
    \label{fig:verify_core_nohw_console}
\end{figure}

\para{Core Logic} We verified that the core logic demonstrates required states in software-only environment, proving the base-level functionality of this phantom aircraft generator. We note that altitude information is as expected. Packet throughput and latency are evaluated as they pertain to maintaining the phantom aircraft's tract in the victim's TCAS with the correct range estimation. However, the computational overhead is too large and imprecise for range estimation, as shown in Figure~\ref{fig:verify_core_nohw_console}. Further investigation into improving the RTT estimation and packet throughput is required.

\section{Discussion and Limitations}
Based on our experience, the implementation of a phantom aircraft generator will require the removal of expensive message interfaces between modules and the implementation of high-accuracy time-stamping in the code.

Heretofore the nature of TCAS, as a highly complex system, has allowed it to enjoy a sort of security through obscurity"~\cite{misc:obscurity}. We have, however, shown that a relatively low-resourced attacker can reproduce the essential signals of TCAS so as to mount an attack against it. Those with the time and resources to attack a complicated system will eventually succeed. It is imperative for defenders to act now in securing the integrity of these systems. 

As future work, it is also essential to develop a platform in which implementations can be tested on real TCAS hardware. TCAS cannot operate independently from its inputs; it has subroutines which disable the TCAS if inputs are failing or nonexistent~\cite{manual:annex10_vol4, manual:acas_manual, manual:intro_tcas_ii}.  Therefore, designing a test bench in which TCAS can operate on real or simulated inputs would be invaluable in conducting accurate security research. Future attack models will focus on a more specific approach and conduct the spoofed message calculation and modulation for expected attack packets beforehand to reduce latency. The degree in which an attacker could control airspace given a fully functional phantom aircraft generator will be also explored. 

\section{Conclusions}
We presented an exploration of TCAS vulnerabilities to exploitation. A wide breadth of analysis is done with respect to TCAS security with the intention to motivate academic and industry-led research into the security of safety-critical airborne collision avoidance systems. 
To accomplish this goal, the following actions are taken. First, a TCAS safety study is analyzed from an adversarial perspective to quantify the effect of attacks on the overall NMAC risk ratio. Second, attacks on TCAS are explored through the model of an attack tree. Third, an attack is chosen, and a threat model is defined to relay the attacker’s goals and capabilities. Finally, an implementation of a threat using GNU Radio is presented, and critical components of the implementation are tested.

\bibliographystyle{IEEEtran}
\bibliography{refs/refs.bib}

\end{document}